\documentclass[12pt]{article}
\usepackage{graphicx}
\usepackage{amsmath,bm}

\newcommand\pubnumber{CIPANP2018-Dudek\\JLAB-THY-18-2813}
\newcommand\pubdate{\today}

\def\jlab{Thomas Jefferson National Accelerator Facility,\\ 12000 Jefferson Avenue, Newport News, VA 23606, USA}
\def\wm{Department of Physics, College of William and Mary,\\ Williamsburg, VA 23187, USA}

\def\support{\footnote{Work supported by the U.S. Department of Energy by contracts DE-AC05-06OR23177, under which Jefferson Science Associates, LLC, manages and operates Jefferson Lab, and DE-SC0018416, which supports Dudek at William and Mary.}}

\def\Title#1{\begin{center} {\Large #1 } \end{center}}
\def\Author#1{\begin{center}{ \sc #1} \end{center}}
\def\Address#1{\begin{center}{ \it #1} \end{center}}

\newcommand\pubblock{\rightline{\begin{tabular}{l} \pubnumber\\
         \pubdate  \end{tabular}}}
\newenvironment{Abstract}{\begin{quotation}  }{\end{quotation}}

\newenvironment{Presented}
{ \begin{quotation} \begin{center}\hspace{-6mm}PRESENTED AT\end{center}\bigskip \begin{center}\begin{large}}
{\end{large}\end{center}\end{quotation}}
      
\def\Acknowledgements{\bigskip  \bigskip \begin{center} \begin{large}
             \bf ACKNOWLEDGEMENTS \end{large}\end{center}}

\begin{document}
\begin{titlepage}
\pubblock

\vfill
\Title{Recent progress on hadron spectroscopy from lattice QCD}
\vfill
\Author{Jozef Dudek\support}
\Address{\jlab \\[3ex] \wm}
\vfill
\begin{Abstract}
\noindent Lattice QCD has matured to a degree where it is now possible to study excited hadrons as they truly appear in nature, as short-lived resonant enhancements decaying into multiple possible final states.\\[-1.5ex]

\noindent Through variational analysis of matrices of correlation functions computed with large bases of interpolating fields it has proven possible to extract many excited state energy levels, and these can be used to constrain the hadron-hadron scattering amplitudes in which hadron resonances can be observed.\\[-1.5ex]

\noindent Recent progress is illustrated with several examples including coupled-channel scattering in the $\pi \eta, K\overline{K}$ and $\pi\pi, K\overline{K}, \eta\eta$ systems in which the $a_0, f_0$ scalar mesons appear.

\end{Abstract}
\vfill
\begin{Presented}
CIPANP2018\\[2ex]
Thirteenth Conference on the Intersections of Particle and Nuclear Physics\\[2ex]
Palm Springs CA, May 29 -- June 3, 2018
\end{Presented}
\vfill
\end{titlepage}
\def\thefootnote{\fnsymbol{footnote}}
\setcounter{footnote}{0}
%

\section{Hadron spectroscopy and QCD}
Quantum Chromodynamics (QCD) is an established part of the standard model of particle physics, being responsible for the confinement of quarks and gluons inside hadrons. Decades of experimental work have uncovered a zoo of hadrons, from the stable or near-stable like the nucleons, pions and kaons, to very short-lived resonances whose existence is inferred from enhancements in the distributions of their strong decay products. Despite having QCD in place to describe this physics, significant mysteries remain in how hadrons are built from quarks and gluons~\cite{Shepherd:2016dni}. 

Examples of this lack of understanding include the recent observation of states such as the `XYZ' family in the charm region~\cite{Lebed:2016hpi}, which were \emph{unexpected} based upon the QCD-based phenomenology developed over many years. In fact this kind of confusion is not only a contemporary problem, having been present even for several long-established states -- as an example, resonances that arguably one might expect to be the some of the simplest, $J^P=0^+$ resonances at low mass, which have only a rather limited set of decay modes, prove to be extremely hard to explain. As well as observations which lack theoretical explanation, there are also cases of theoretical expectations not being unambiguously observed in experiment. Examples include states in which an excitation of the gluonic field plays a role, either alone in a `glueball', or coupled to quarks in a `hybrid'~\cite{Dudek:2012ag, Dudek:2011bn, Meyer:2015eta}.

A significant complicating factor in the study of hadron spectroscopy is that the vast majority of hadrons are short-lived resonances, with the same nonperturbative QCD dynamics responsible for their decay as is in charge of binding the states. This demands that we consider an approach to QCD which is capable of dealing with all this physics on the same footing, and a first-principles approach in which this is possible is provided by \emph{lattice QCD}.

\section{The lattice as a tool for QCD}

The most successful approach to calculating QCD nonperturbatively is to numerically compute the theory discretized on a finite grid of space-time points. By Monte-Carlo sampling possible gluon field configurations according to a probability distribution dictated by the lagrangian of QCD, correlation functions can be estimated via a path-integral average over the configurations. This is a controlled approximation to QCD in that the introduced features, the nonzero lattice spacing and the finite space-time volume, can be adjusted, and the behavior of observables as a function of them studied, and the appropriate continuum and infinite-volume limits established. In many practical calculations, another adjustment to the theory is made, by computing with quark mass parameter values somewhat larger than the very small masses inferred from experimental data. While the principal reasoning behind this is to reduce the computational cost by dealing with better conditioned and smaller matrices, a secondary reason will reveal itself later, related to the lightness of the pion allowing many decay channels to be open for resonant states.

The field of lattice QCD has matured~\cite{Kronfeld:2012uk} to stage where there is sufficient systematic control over the approximations made, that calculations of hadrons stable under the strong interaction can be considered to be \emph{precise},  with the possibility, in certain observables, of using any deviation between experiment and lattice QCD as a signal of physics going beyond the standard model~\cite{Aoki:2016frl}. These stable states are energy eigenstates of QCD, typically the lightest with a given set of quantum numbers, and their properties can be extracted from lattice QCD correlation functions having large time separations between operators. For the case of unstable resonances, the situation is not as simple: they are not eigenstates of definite energy, rather they appear as enhancements in a continuum of scattering states. In order to understand the methodology by which they are studied in lattice QCD, one must understand how this continuum maps onto the \emph{discrete} spectrum of states present in the finite spatial volume defined by the lattice.

\section{`Scattering' in a finite-volume}

The essential physics can be illustrated using a simple elastic scattering system in one-dimensional quantum mechanics~\cite{DeGrand:2006zz}. If we consider two identical bosons moving along a line with a finite-range potential between them, we know that outside the range of the potential, the wavefunction is just that of the free system with a momentum dependent phase-shift, $\psi(x) \sim \cos \big( |k| x + \delta(k) \big)$. The phase-shift contains all the information one can have about the scattering, and it can be obtained by matching the solution of the Schr\"odinger equation inside the potential to the form above at the range of the potential. A resonance would appear as a rapid change in value of the phase-shift with continuous variation of the momentum, $k$.

We now consider our one-dimensional system as being of fixed length $L$ with a periodic boundary condition (i.e. the line has become a circle). Insisting that the wavefunction and its derivative are continuous at the periodic boundary leads to a quantization condition on the momentum, $k = \frac{2\pi}{L} n - \frac{2}{L} \delta(k)$, for integer values of $n$. In this way we see that the finite-volume system has a discrete spectrum which can be related to the scattering amplitude of the infinite-volume system. This spectrum will vary in a predictable way with changing volume if the scattering amplitude is known. The application of this logic to lattice QCD will be the reverse: the discrete spectrum in one or more volumes is calculated, and from it a discrete set of points on the phase-shift curve are determined.

A more sophisticated derivation relating the discrete spectrum to infinite-volume elastic scattering amplitudes within quantum field theory in three spatial dimensions leads to a result that is conceptually the same, with the main differences being technical in nature due to the mismatch between the cubic symmetry of the periodic boundary and the continuous rotational symmetry one assumes in the construction of partial-waves~\cite{Luscher:1986pf}. A complete list of references describing extensions of the formalism to describe moving frames, coupled-channel scattering and more can be found in a recent review, Ref.~\cite{Briceno:2017max}.

\section{The $\rho\,$-resonance in elastic $\pi\pi$ scattering}

The $J^P=1^-$ isospin-1 $\rho$ resonance which appears in $P$-wave $\pi\pi$ scattering is a useful test-bed for scattering calculations in lattice QCD. The state appears as a resonance even for heavier than physical light quark masses, provided the pion mass remains below about 400 MeV, where it becomes a stable bound-state. A significant number of lattice QCD calculations which consider this channel have now been carried out (for a citation list see the review, Ref.~\cite{Briceno:2017max}).

An approach which has proven very successful is to compute a \emph{matrix of two-point correlation functions} using a basis of operators at source and sink. The operators, constructed out of quark and gluon fields, which prove to be most effective in interpolating the relevant states from the vacuum include fermion bilinears with vector quantum numbers (which resemble $q\bar{q}$-like constructions) supplemented with meson-meson-like constructions built from products of pseudoscalar fermion-bilinears each of definite momentum.

Each element of the matrix of two-point functions in principle receives a contribution proportional to $e^{-E_n t}$ from every QCD eigenstate $|n\rangle$ of energy $E_n$, with the appropriate quantum numbers, but with a different weight, and the use of a \emph{variational} solution method, which essentially diagonalizes the matrix, leads to determination of many states. Essentially, each state in the spectrum is proposed to be interpolated optimally by a particular linear combination of the basis operators, with each different state having an orthogonal combination. This use of linear algebra allows us to deal with cases in which states appear that are almost degenerate in energy, and providing the operator basis is large and diverse enough, there is essentially no restriction on the number of energy states that can be extracted.

After this variational analysis, with the discrete energies in hand, discrete values of the phase-shift at the relevant energies can be found using methods analogous to those described in the previous section. Early attempts to study $\pi\pi$ scattering (e.g. the pioneering work~\cite{Aoki:2007rd}) used relatively small operator bases in a single volume, allowing determination of only a few points on the phase-shift curve, which were nevertheless compatible with the presence of a resonance.

\begin{figure}[tb]
\centering
\includegraphics[width=2.9in]{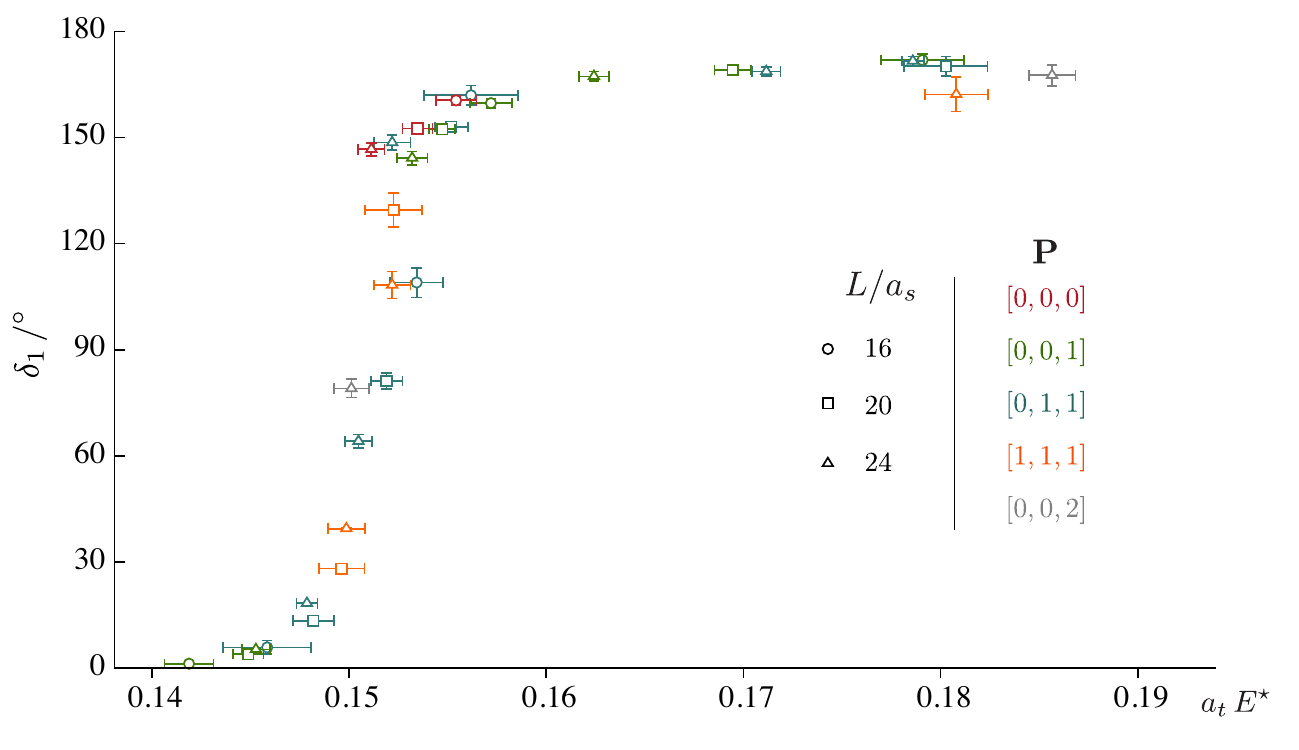}
\includegraphics[width=2.9in]{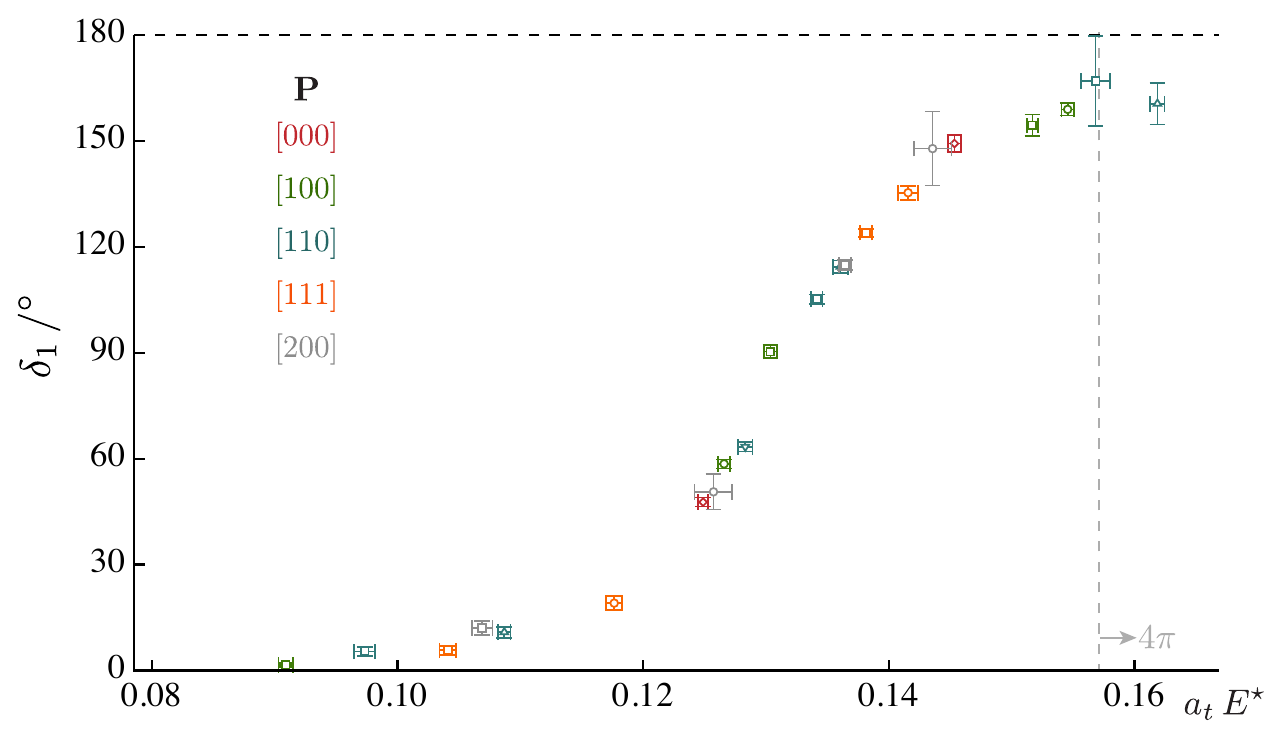}
\caption{$\pi\pi$ $P$-wave elastic scattering phase-shift determined for $m_\pi=391$ MeV~\cite{Dudek:2012xn} (left, three spatial volumes: $16^3, 20^3, 24^3$), and $m_\pi=236$ MeV~\cite{Wilson:2015dqa} (right, single volume: $32^3$).}
\label{fig:rho}
\end{figure}

Examples of contemporary calculations~\cite{Dudek:2012xn, Wilson:2015dqa} are shown in Figure~\ref{fig:rho}. These works made use of a large basis of operators, several moving frames, and either multiple volumes~\cite{Dudek:2012xn}, or a single large volume in which the density of states is high~\cite{Wilson:2015dqa}. We see that it is possible to map out the phase-shift curve in detail, leading to an unambiguous signal for a narrow resonance corresponding to the $\rho$. The evolution of the mass and width of the resonance with changing light quark mass is much as one would expect -- the mass decreases and the quark mass decreases while the width into $\pi\pi$ increases owing to the larger phase-space.

\section{Coupled-channel scattering}

While the $\rho$ resonance in elastic $\pi\pi$ scattering provides a useful test-bed for lattice QCD techniques, it is a rather unique system when hadron spectroscopy is viewed globally -- most hadrons decay into more than one final state, that is to say that they are resonances in \emph{coupled-channel} scattering, where a \emph{matrix} in the space of kinematically open channels is required to describe scattering.

In a finite spatial volume, $L \times L \times L$, the quantization condition reads
\begin{equation}
\det \big[ \bm{1} + i \bm{\rho}(E) \, \bm{t}(E) \big( \bm{1} + i \bm{\mathcal{M}}(E,L) \big) \big] = 0, \label{qc}
\end{equation}
where $\bm{t}(E)$ is the coupled-channel scattering $t$-matrix, $\bm{\rho}(E)$ is the relativistic phase-space, and $\bm{\mathcal{M}}(E,L)$ is a matrix of known kinematic functions encoding the finite-volume nature of the problem. The determinant is taken over the space of kinematically open scattering channels
\footnote{and also over the different partial-waves \emph{subduced} into the particular irreducible representation of the lattice symmetry group under consideration, but let us ignore that complication here.
}
. Using this expression to determine coupled-channel scattering amplitudes given a finite-volume spectrum is a qualitatively different problem than that present in the elastic case, as per energy level it is a function of multiple unknown entries in the $t$-matrix.

One approach which has proven successful is to parameterize the energy-dependence of the $t$-matrix in a manner that respects coupled-channel unitarity, and by varying the free parameters, attempt to obtain a best global description of the complete finite-volume spectra computed in one or more frames and in one or more volumes~\cite{Guo:2012hv}. 

The first explicit application of this technique was to the problem of coupled $\pi K, \eta K$ scattering in $J^P=0^+, 1^-, 2^+$ partial-waves~\cite{Dudek:2014qha, Wilson:2014cna}. Using spectra computed with \mbox{391 MeV} pions in three volumes in a number of moving frames, the scattering amplitudes could be determined with some confidence, and there proved to be relatively little coupling between the two channels, with the $\pi K$ partial-wave amplitudes showing a range of behaviors including a narrow resonance in the $D$-wave and a stable bound state in the $P$-wave as well as more unusual structures in $S$-wave. 

In the past couple of years the Hadron Spectrum Collaboration has reported calculations which show meson-meson scattering channels which couple strongly, and these calculations offer new insight into the longstanding problem of understanding the light scalar meson resonances, the $f_0(980)$ and the $a_0(980)$.

\subsection{Coupled $\pi\eta, K\overline{K}$ scattering and the $a_0$ meson}

\begin{figure}[tb]
\centering
\includegraphics[width=5in]{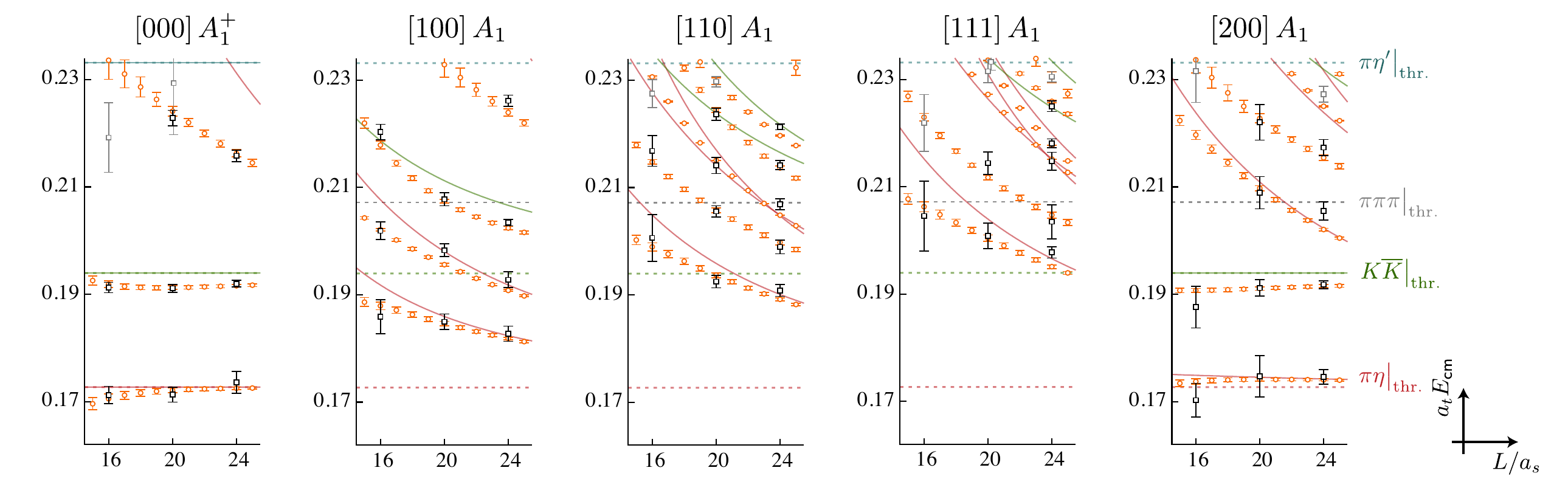}
\includegraphics[width=3.5in]{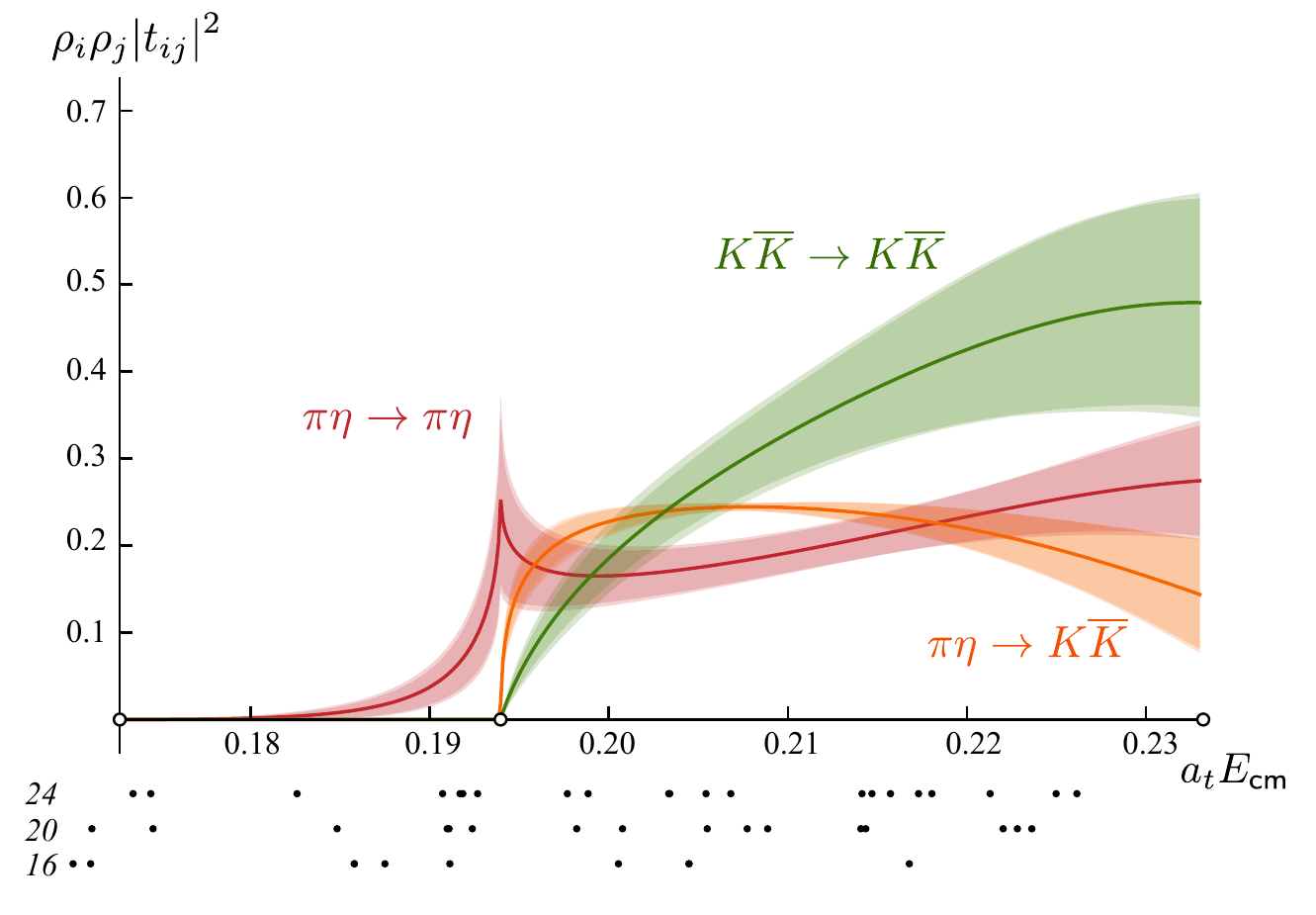}
\caption{Taken from Ref.~\cite{Dudek:2016cru}. (top) The lattice QCD finite volume spectrum in three volumes in the rest frame and four moving frames (black points) together with the spectrum predicted by Eqn.~\ref{qc} for a particular parameterization of the coupled $\pi\eta, K\overline{K}$ $t$-matrix (bottom). }
\label{fig:a0}
\end{figure}

Experimentally the $a_0(980)$ is a well-established enhancement which typically appears as a peak very close to the $K\overline{K}$ threshold in processes leading to a $\pi \eta$ final state in $S$-wave. In Ref.~\cite{Dudek:2016cru}, the Hadron Spectrum Collaboration computed finite volume spectra with the relevant quantum numbers to contain the coupled $\pi \eta, K\overline{K}$ scattering system, in three volumes and multiple moving frames. The calculations were done with light quarks that were heavier than they are in reality such that the pions were of mass 391 MeV, so the results are not directly comparable with experiment. The energy levels obtained spanned the energy region from slightly below $\pi \eta$ threshold up to and beyond the $\pi \eta'$ threshold, including a significant region of energy in which coupled $\pi \eta, K\overline{K}$ scattering is kinematically allowed. In total 47 energy levels were used to constrain parameterizations of the $J^P=0^+$ amplitude -- an example of one such parameterization featuring six free parameters is shown in Figure~\ref{fig:a0}.

A clear sharp enhancement is visible in the $\pi \eta \to \pi \eta$ amplitude at the $K\overline{K}$ threshold, which is observed to occur together with a rapid turning-on of the $K\overline{K}$ final state. This behavior is not the canonical appearance of a coupled-channel resonance, and in principle, a cusp can occur at the opening of each new channel, so while this enhancement is suggestive, a more robust justification is required if we wish to claim a resonance is present. 

\begin{figure}[tb]
\centering
\includegraphics[width=3.7in]{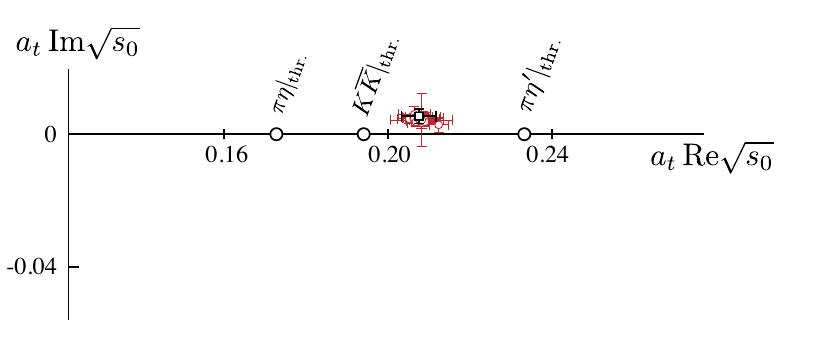}
\includegraphics[width=2in]{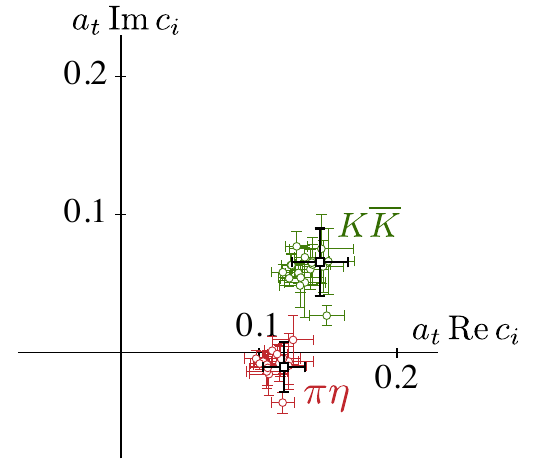}
\caption{Taken from Ref.~\cite{Dudek:2016cru}. (left) Red points show the pole in the complex energy plane found for each of a large number of $t$-matrix parameterizations. (right) The couplings which follow from factorizing the residue at the pole, $t_{ij} \sim \frac{c_i \cdot c_j}{s_0 -s}$. }
\label{fig:a0res}
\end{figure}

Fortunately, there is a well established technique for determining the resonance content of scattering amplitudes: analytic continuation into the complex energy plane. The idea is that resonances should appear as \emph{pole singularities} at complex values of the scattering energy, with the real part of the pole position often identified with the mass, and the imaginary part with the width. The residue of the pole in each element of the $t$-matrix can be factorized into a product of numbers which can be interpreted as \emph{couplings} of the resonance to the various decay channels
\footnote{In fact the analytic continuation into the complex plane is complicated by the multi-sheeted nature of the coupled-channel system, but we will not discuss this subtlety here.}
. Figure~\ref{fig:a0res} shows that indeed there is a resonance pole present, lying very close to the $K\overline{K}$ threshold, and the coupling of this resonance to the $K\overline{K}$ channel is found to be at least as large as its coupling to the $\pi \eta$ channel, explaining the rapid turn-on of the $K\overline{K}$ final-state.

In order to ensure that this result is not a quirk of the particular parameterization choice made, a large number of parameterization variations were explored, and very little variation was observed in the energy dependence of the amplitudes in the region where energy levels provide constraint. Furthermore the resonance pole is present in the same region of complex energy for every parameterization capable of describing the finite-volume spectrum. Taking the small variation in the pole position and couplings over parameterization into account in estimation of a conservative `systematic' error, this calculation at $m_\pi = 391$ MeV finds that an $a_0$-like resonance is present with $m = 1177(27)$ MeV, $\Gamma = 49(33)$ MeV, and with a ratio of coupling constants $\left| c(K\overline{K}) / c(\eta \pi) \right|^2 = 1.7(6)$.

\subsection{Coupled $\pi\pi, K\overline{K}, \eta\eta$ scattering and $f_0$ mesons}

Experimentally the isoscalar channel with $J^P=0^+$ is rather complicated --  a very broad bump which can be traced to the $\sigma$ resonance is the backdrop to the $f_0(980)$ which appears as a sharp dip in the $\pi\pi \to \pi\pi$ amplitude at the $K\overline{K}$ threshold. Along with this dip, there is a rapid turn-on of the $K\overline{K}$ final state as its threshold opens.

\begin{figure}[tb]
\centering
\includegraphics[height=2in]{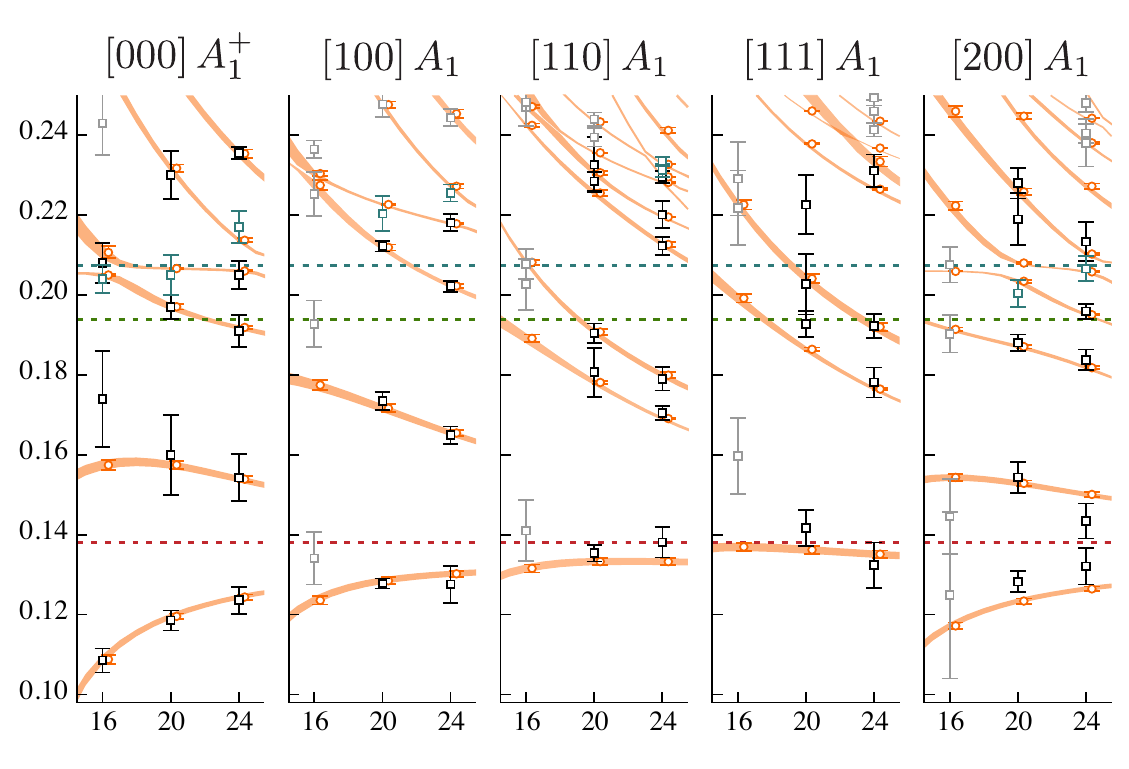}
\includegraphics[width=3in]{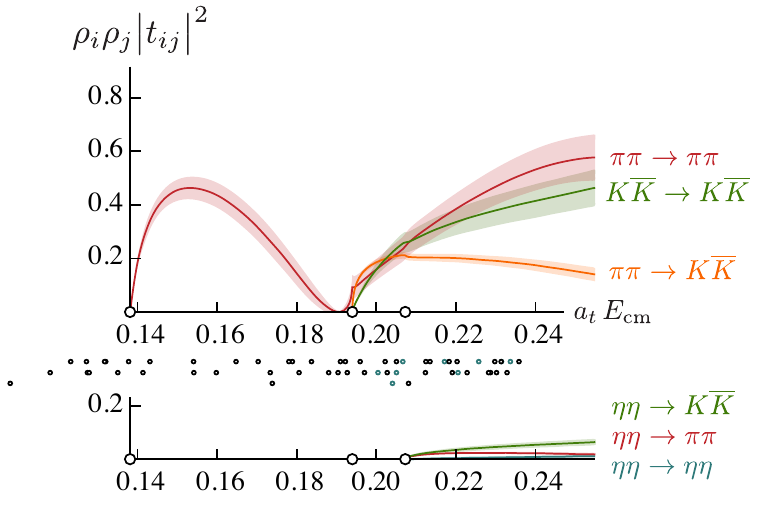}
\caption{Taken from Ref.~\cite{Briceno:2017qmb}. (left) The lattice QCD finite volume spectrum in three volumes in the rest frame and four moving frames (black/blue points) together with the spectrum predicted by Eqn.~\ref{qc} for a particular parameterization of the coupled $\pi\pi, K\overline{K}, \eta\eta$ $t$-matrix (right). }
\label{fig:f0}
\end{figure}

A calculation of this system, using the same lattices as the $a_0$ calculation above, was reported on in Ref.~\cite{Briceno:2017qmb}. With the quark masses used, there is a very limited energy region of coupled two-channel $\pi\pi, K\overline{K}$ scattering, as the $\eta \eta$ channel opens up less than 100 MeV above $K\overline{K}$, meaning that the calculation was required to consider \emph{three-channel} scattering. 57 energy levels across three volumes were used to constrain parameterizations of the three-channel $t$-matrix, with one example parameterization, which featured eight free parameters, being shown in Figure~\ref{fig:f0}.

Three features are immediately clear: the rapid turn-on of the $\pi\pi$ amplitude at threshold (in distinction to the slow growth in $\pi \eta$ in the previous calculation), the dip in $\pi\pi$ just before the $K\overline{K}$ threshold that appears with a rapid turn-on of the $K\overline{K}$ final state, and finally the lack of any significant activity in the $\eta \eta$ channel. 

All these features can be explained in terms of the singularity content of this amplitude. The large effect at $\pi\pi$ threshold proves to be due to a bound-state lying just below threshold that we can associate with the $\sigma$ at this larger than physical pion mass. The dip at $K\overline{K}$ threshold coupled with the rapid rise in the $K\overline{K}$ amplitudes is traced to a resonance pole out in the complex energy plane lying close to the $K\overline{K}$ threshold. From the residues of this pole, we conclude that this resonance is strongly coupled to both $\pi\pi$ and $K\overline{K}$ channels, but only weakly coupled to $\eta\eta$. These two singularities, the bound-state and the resonance, are found to be present in all amplitude parameterizations capable of describing the finite-volume spectra, with best estimates being that the stable $\sigma$ has a mass of $745(5)$ MeV, while the $f_0$-like resonance has $m = 1166(45)$ MeV, $\Gamma = 181(68)$ MeV, and $\left| c(K\overline{K}) / c(\pi \pi) \right|^2 = 1.4(6)$.

It is worth noting that while the energy dependencies shown in Figures~\ref{fig:a0} and~\ref{fig:f0} are superficially rather different (sharp asymmetric peak versus broad dip), they prove to be due to resonances ($a_0$-like and $f_0$-like) that are appear to be closely related -- they have a similar mass and similar coupling strengths to the $K\overline{K}$ final state, and the main difference between them, the larger width of the $f_0$-like state, can likely be traced to the larger phase-space for decay into two light pions.

\begin{figure}[tb]
\centering
\includegraphics[width=2.85in]{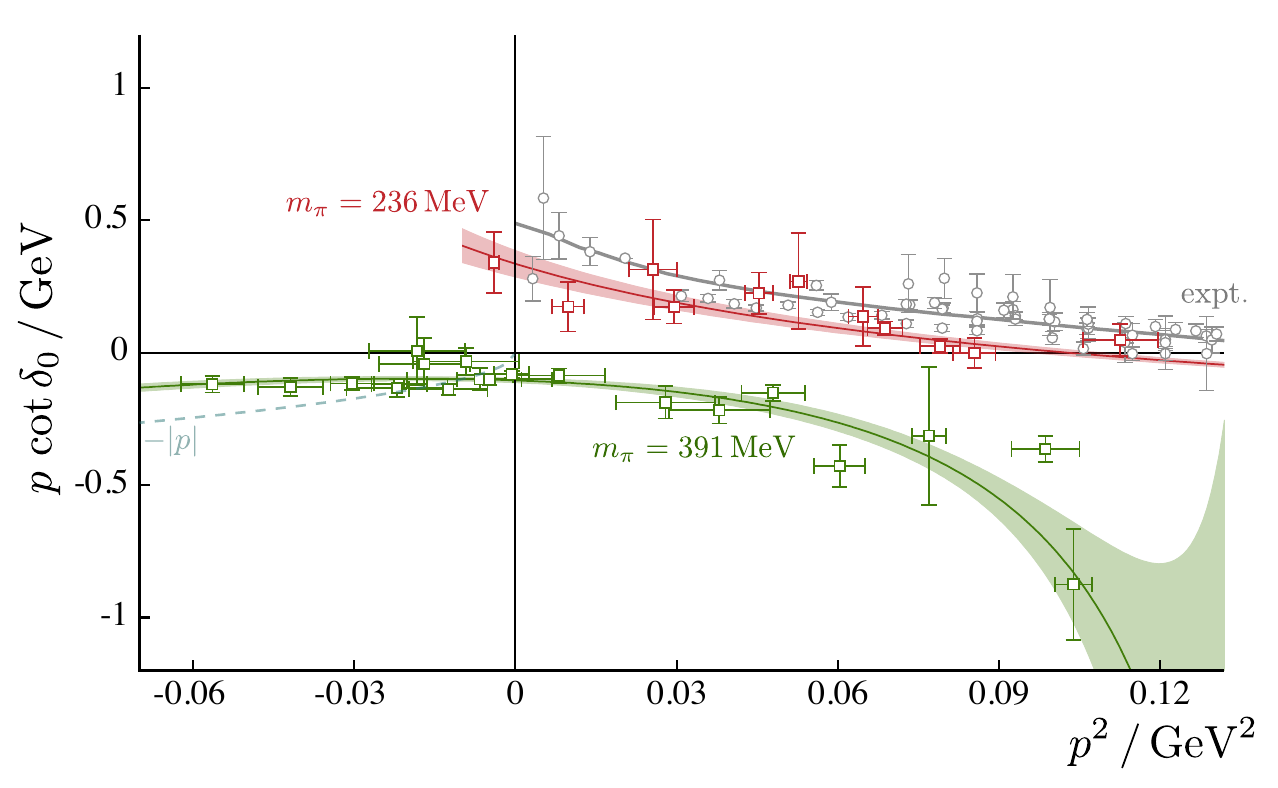}
\includegraphics[width=2.85in]{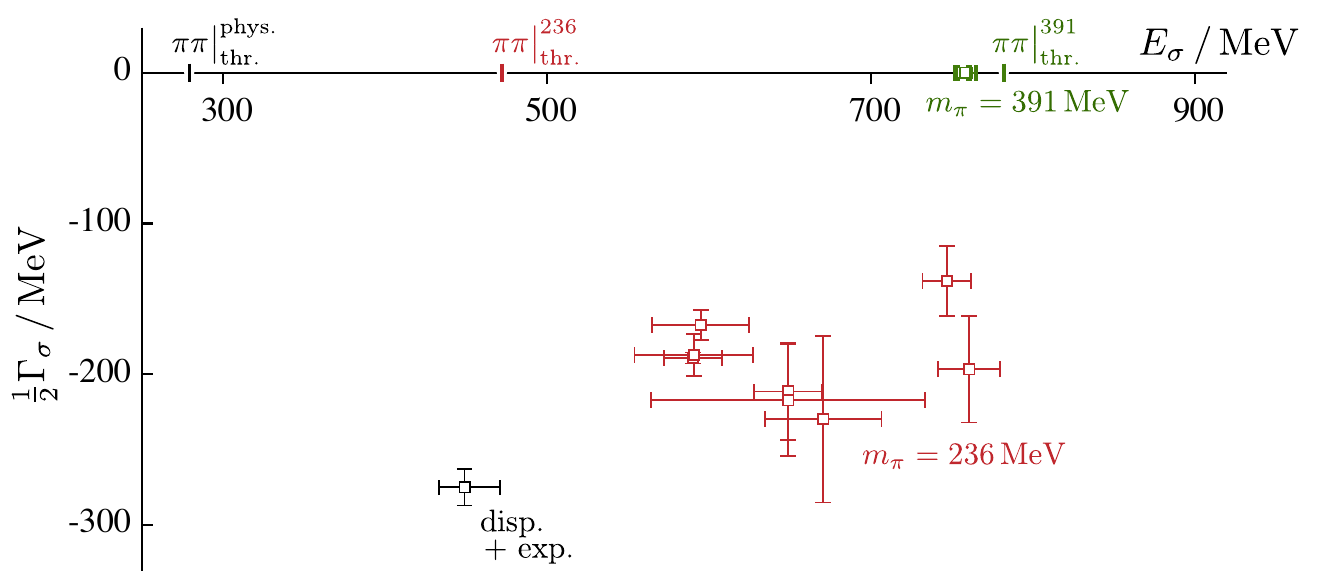}
\caption{Elastic $\pi\pi$ scattering in $S$-wave determined at two pion masses~\cite{Briceno:2016mjc}. (left) Phase-shift. (right) Pole singularities for a variety of parameterization forms. }
\label{fig:sigma}
\end{figure}

Of course, as the light quark mass is reduced towards the physical value, and the pions approach 140 MeV, we expect the $\sigma$ to evolve into a broad resonance. Behavior in this direction was explicitly observed in a calculation of \emph{elastic} $\pi \pi$ scattering with $m_\pi = 236$ MeV reported on in Ref.~\cite{Briceno:2016mjc}, and shown here in Figure~\ref{fig:sigma}, where the qualitative change in the behavior of the scattering phase-shift can be traced back to the $\sigma$ pole moving off the real axis into the complex plane. A similar calculation in Ref.~\cite{Guo:2018zss}, analyzed in context of unitarized chiral effective theory, found a compatible result.

\section{Coupling resonances to external currents}

The methods described in the previous sections offer the opportunity to determine the presence of resonances within a first-principles approach to QCD. However they offer only limited insight into the structure of the states -- valuable information in this regard comes if we consider the coupling of resonances to external currents, which may probe the distribution of, for example, charge within the states. 

The formalism is in place~\cite{Briceno:2014uqa} to account for the effect of working in a finite-volume when considering \emph{transition matrix elements} where a current induces a transformation from a stable hadron into a hadron-hadron final state, or when the current itself fluctuates into a hadron-hadron pair. An application of the latter case is the process $\gamma^\star \to \pi\pi$, which can equivalently be thought of as the timelike vector form-factor of the pion. As presented in Ref.~\cite{Feng:2014gba}, the contribution of the $\rho$ resonance can be clearly observed in the energy-dependence of this quantity when computed in lattice QCD.

A more complicated case is $\gamma^\star \pi \to \pi \pi$ where the amplitude is now a function of two kinematic variables: the $\pi\pi$ energy and the photon virtuality, $Q^2$. A calculation presented in Ref.~\cite{Briceno:2016kkp, Briceno:2015dca} showed the presence of the resonant $\rho$, and by parameterizing the amplitude, it was possible to analytically continue to the $\rho$ pole and from the residue of the pole extract the $\rho \to \pi \gamma^\star$ transition from-factor as a function of $Q^2$. 

One can imagine that future applications of these methods might study form-factors of somewhat more mysterious resonances like the light scalars, and provide information about their internal structure. 

\section{Opportunities and Challenges}

Studying properties of hadron resonances by taking advantage of the finite-volume in lattice QCD has become a practical reality~\cite{Briceno:2017max}. The methodology utilized in the calculations described above which considered low-lying resonances like the $\rho$, the $\sigma$ and the $f_0, a_0$, can now be extended to consider more challenging cases.

The `XYZ' states in the charmonium region generally lie in energy regions in which several decay channels are kinematically accessible. The finite-volume approach described above has one very important difference with respect to the kind of analysis typically followed in experiment: individual decay channels cannot be considered in isolation, all kinematically open channels influence the finite-volume energies simultaneously, and the entire $t$-matrix must be considered at once. Complications in this sector come not only from the number of open channels, but also from the fact that the scattering hadrons involved are not all spinless, e.g. the $D^*$ or the $J/\psi$ may feature
\footnote{both of which can be considered in many calculations to be stable without introducing significant error}
. The formalism to deal with scattering of spinning hadrons is in place~\cite{Briceno:2014oea}, and has recently been tested in a calculation of $\pi \rho$ scattering at a large value of the light quark mass such that the $\rho$ becomes stable~\cite{Woss:2018irj}. The `$Z_c$' states observed in final states like $J/\psi \, \pi$ have been argued to be \emph{tetraquark} candidates (i.e. having internal structure $cu \bar{c}\bar{d}$), which suggests that we might consider augmenting the lattice QCD operator basis to include local four-quark constructions. This was explored in detail in Ref.~\cite{Cheung:2017tnt}, where it was found that in the channels where the $Z_c$ states are expected to be found, adding a basis of tetraquark operators in addition to the complete expected set of low-energy meson-meson operators led to no significant change in the spectrum, and the spectrum extracted showed no clear sign of narrow resonance behavior. Further calculations which extract the coupled-channel scattering amplitudes are warranted, and are underway.

The search for hybrid hadrons, states in which the gluonic field plays a vital role, continues in experimental programs at COMPASS, GlueX and CLAS12. Simplified lattice QCD calculations in which the decay properties of excited hadrons are ignored~\cite{Dudek:2012ag,Dudek:2011bn,Dudek:2013yja,Dudek:2011tt,Dudek:2010wm,Dudek:2009qf} seem to suggest that a spectrum of hybrid mesons and baryons should be expected with a characteristic unique set of spin-parity quantum numbers. The next generation of lattice QCD calculations will need to relax the uncontrolled approximation of assuming they are stable, and establish these states as resonances with couplings to open decay channels. This can be done in a somewhat simplified setting initially by working with rather heavy light quarks, such that the number of open decay channels is manageable with current techniques.

The ability to study the light scalar mesons within lattice QCD with their appearance being as it is experimentally, as coupled-channel resonances, opens up a new hope for finally understanding their origin. A first target is to track the behavior of the resonance poles with changing quark mass, in particular to determine if the $f_0, a_0$ resonances are tied to the $K\overline{K}$ threshold, as the combination of experimental data and the calculations with $m_\pi = 391$ MeV would seem to suggest. Calculations coupling these resonances to external currents will allow their transition form-factors to be determined, and using this additional information, which will offer suggestions as to the physical size of the states, it may be possible to infer the importance of `spatially large' meson-meson molecular configurations relative to `spatially small' confined $q\bar{q}$ arrangements.

A challenge which faces the field is to consider cases in which three-body final states become kinematically accessible. There is, as yet, no complete formalism which relates three-body scattering amplitudes to the discrete spectrum of states in a finite volume, but there is much activity in this direction. A practical formalism to deal with three-body and higher configurations becomes particularly vital if we imagine reducing the pion mass toward the physical value, where virtually all resonances lie above multi-hadron thresholds.

The progress seen recently in lattice QCD calculations, which makes use of the finite spatial volume of the lattice to extract information about scattering amplitudes, suggests that we may be moving into an era in which understanding about excited hadron spectroscopy can come directly from QCD.

\Acknowledgements
Much of the work presented above was performed by the Hadron Spectrum Collaboration ({\tt www.hadspec.org}). I wish to offer particular thanks to my coauthors on the papers discussed in some detail, Raul Briceno, Robert Edwards, and David Wilson.\\

\noindent The conference presentation was dedicated to the memory of Michael Pennington. Mike was an unfailing supporter of our efforts to study hadron spectroscopy using lattice QCD tools. His advocacy, constructive criticism, and enthusiasm as Jefferson Lab Theory Center leader were significant elements in our recent progress. Mike was much more than just a good `boss', he was a personal mentor, and a kind friend, and like many others, I have much to thank him for, and I will miss him terribly.

\end{document}